\documentclass[fleqn,twoside]{article}
\usepackage{espcrc2}
\usepackage{graphicx,amsmath}
\usepackage[figuresright]{rotating}

\title{Testing CP and Time Reversal Symmetries with $\Lambda_b \to \Lambda V(1-)$ Decays}

\author{O.~Leitner\address{LPNHE, Groupe Th\'eorie, Universit\'e P. $\&$ M. Curie,
                    4, Place Jussieu, F-75252, PARIS.}
          and
        Z.~J.~Ajaltouni\address{LPC/IN2P3-CNRS, Universit\'e Blaise Pascal, 
        F-63177, AUBIERE.}}

\begin{document}

%-----------------------------------------------------------------------------
\begin{abstract}
In this letter, an overview is given for interesting tests of both CP and 
Time Reversal symmetries with the beauty baryon $\Lambda_b $. Extensive use 
of the helicity formalism and HQET is done for all calculations.  
Then, emphasis is put on sophisticated methods like analysis of 
resonance polarizations and particular angle distributions 
which can exhibit a clear signal of TR violation.
\end{abstract}
%-----------------------------------------------------------------------------

\maketitle

%-----------------------------------------------------------------------------
\section{PHYSICAL MOTIVATIONS}
Modern Particle Physics is well described by gauge field theories which are built 
on fundamental principles, the most important ones being Lorentz Invariance, 
Unitarity and Hermiticity. An immediate consequence of these principles is the 
famous CPT theorem which stipulates that any physical system and its 
CPT-conjugate one have identical physical laws. On another side, CPT theorem 
supposes that, if CP symmetry is violated, Time Reversal (TR) is no 
longer a good symmetry and this mathematical feature represents only an {\it 
indirect violation} of TR symmetry. However some experiments performed recently 
at CERN and Fermilab have shown a clear signal of {\it direct} TR violation in 
the $K^0 \bar K^0 $ system. Our aim is to demonstrate that search for direct 
TR violation can be performed at the LHC energies by studying the cascade decays 
of $\Lambda_b \to \Lambda V, \  \ \Lambda \to  p \pi , \  V  \to  x^+ x^-$, $x$ 
being a lepton or a pion. Our main tool is constructing {\it Triple Product 
Correlations} (TPC) defined by the following relations
 ${\vec {v_i}} =  \vec {p_i} ,  \vec{s_i} \ , \ C_{ijk} = {\vec {v_i}} 
\cdot {({\vec {v_j}} \times {\vec {v_k}})} \rightarrow TR \to  - C_{ijk}$
 which are odd by TR. A good example of TPC is the Transverse Polarization 
of a spinning resonance coming from $\Lambda_b$ decays.        
%%%%%%%%%%%%%%%%%%%%%%%%%%%%%%%%%%%%%%%%%%%%%%%%%%%%%%%%%%%%%%%%%%%%%
\begin{table*}[hpt]
\caption{Branching ratio, $\mathcal{BR}$, for $\Lambda_b \to \Lambda J/\Psi, 
\Lambda_b \to \Lambda \rho^0$ and  $\Lambda_b \to \Lambda \omega$.}
\label{table:1}
\newcommand{\m}{\hphantom{$-$}}
\newcommand{\cc}[1]{\multicolumn{1}{c}{#1}}
\renewcommand{\tabcolsep}{1.7pc} % enlarge column spacing
\renewcommand{\arraystretch}{1.2} % enlarge line spacing
\begin{tabular}{@{}|l|c|c|c|c|}
 \hline
 $ N^{eff}_c$&2&2.5&3&3.5\\
\hline
 ${\Lambda} J/{\psi}$&$8.95 \times {10^{-4}}$&$2.79 \times {10^{-4}}$&$0.62 \times
{10^{-4}}$&$0.03 \times {10^{-4}}$\\
\hline
${\Lambda} {\rho}^0$&$1.62 \times {10^{-7}}$&$1.89 \times {10^{-7}}$&$2.2 \times
{10^{-7}}$&$2.4 \times {10^{-7}}$\\
\hline
${\Lambda} {\omega}$&$22.3 \times {10^{-7}}$&$4.75 \times {10^{-7}}$&$0.2 \times
{10^{-7}}$&$0.64 \times {10^{-7}}$\\
\hline
\end{tabular}\\[2pt]
\end{table*}
%%%%%%%%%%%%%%%%%%%%%%%%%%%%%%%%%%%%%%%%%%%%%%%%%%%%%%%%%%%%%%%%%%%%%%%
%------------------------------------------------------------------------------
\section{DYNAMICS OF  $\Lambda_b$ DECAYS}
At LHC energies, $10\%$ of the produced $b \bar b$ pairs hadronize into beauty baryons
  ${\cal B}_b =   {\Lambda_b},  {\Sigma}_b,  {\Xi}_b, ...$ and approximately  $90\%$ of 
${\cal B}_b$ are dominated by $\Lambda_b$. Recently many authors have pointed out the 
possibility of both testing TR invariance and searching for New Physics in
$\Lambda_b$ three-body decays like ${\Lambda}_b  \to  Baryon  \ {\ell}^+ {\ell}^- ,   
Baryon   \ h^+ h^-$, where ${\ell}^{\pm}$ and $ h^{\pm}$ could originate from continuum 
and/or resonances like $J/{\psi}$. The model developed in references~\cite{Ajalt,Leitner:2006nb}  is based
on the same final states but, emphasis is put on measuring {\it physical observables} 
built from the two intermediate resonances, $\Lambda$  and    $J/{\psi}$ or 
${\rho}^0-{\omega}$  which come from $\Lambda_b$ decays. Our method offers some 
advantages: (i) because of $\Lambda_b$ weak decays, the intermediate
resonances are polarized and some components of their vector-polarization  
$\vec {\cal P}$ 
are not invariant by TR; (ii) it is expected that $\Lambda_b$ produced in p-p 
collisions is {\it transversally polarized} like ordinary hyperons in hadron 
collisions and, this physical property will be exploited in order to find many 
interesting TPC parameters which are T-odd, see Ref.~\cite{Leitner:2006nb}. 
%-----------------------------------------------------------------------------
\subsection{Cascade decays}
\noindent
The initial laboratory frame to which $\Lambda_b$ polarization is referred is 
defined like: 
$\vec {e_1} = {\vec {p_1}}/{p_1},     \vec {e_3}=  \frac {\vec{p}_1 
\times \vec{p}_b} {|{\vec{p}_1 \times \vec{p}_b}|}, \  \vec {e_2} = 
{\vec {e_3}} \times {\vec {e_1}}$
 where $\vec {p_1}$ and   $\vec{p}_b$ are respectively the incident 
proton momentum and the produced $\Lambda_b$ one.
 \newline
$\Lambda_b$ being transversally polarized, its polarization value is given 
by ${\cal P}^{\Lambda_b}  =  \langle {\vec {S_{\Lambda_b}}} \cdot {\vec {e_3}} 
\rangle$. Let $M_i$ be 
the $\Lambda_b$ spin projection along $\vec {e_3}$ axis and, $\lambda_1$  and 
$\lambda_2$ are respectively the helicity values of $\Lambda$ and  $V$. 
Conservation of total angular momentum leads to {\it four} possible values 
for the pair $(\lambda_1 , \lambda_2)=(1/2,0), (1/2,1), (-1/2,-1),  
(-1/2,0)$ with $M_i = \pm 1/2,    M_f =  \lambda_1 - \lambda_2   
=  \pm 1/2$. In the following, the Jacob-Wick-Jackson helicity formalism 
will be associated with the method of successive or cascade decays. We define (i) 
${\rho}^{\Lambda_b}$ the polarization density matrix (PDM) of $\Lambda_b$,
(ii) $A_0{(M_i)} = \mathcal{M}_{\Lambda_b}{(\lambda_1 ,\lambda_2)} 
D_{M_i M_f}^{{1/2} \star}{(\phi,\theta,0)}$  the $\Lambda_b$ decay amplitude, 
(iii) $A_1(\lambda_1) \ \mathrm{and} \ A_2(\lambda_2) $ the resonance decay 
amplitudes. So, the total decay amplitude is expressed by:
\begin{eqnarray}
{\cal A}_I = \sum_{\lambda_1,\lambda_2}A_0(M_i) A_1(\lambda_1) A_2(\lambda_2)\ , \nonumber
\end{eqnarray} 
from which the decay probability can be deduced,
\begin{eqnarray}
d\sigma  \propto   \sum_{M_i,M^{\prime}_i}{\rho}_{M_i M^{\prime}_i}^{\Lambda_b}\ . 
{\cal A}_I {\cal A}_I^{*}\ . \nonumber
\end{eqnarray} 
%---------------------------------------------------------------------------------
\subsection{Hadronic matrix element (HME)}
\noindent
Full computation of the HME has been performed by using the 
Operator Product Expansion (OPE) formalism supplemented by 
hypothesis derived from the Heavy Quark Effective Theory (HQET). In the framework 
of the factorization hypothesis, the HME can be written as: 
HME = current products $ \otimes  $ form factors, where (i) form factors 
are computed according to an expansion with $  {\cal O}(1/{m_b}) $ and, (ii) the 
currents are evaluated from the effective hamiltonian, $\mathcal{H}^{eff}=
\frac{G_{F}}{\sqrt{2}} V_{qb}V^{\star}_{qs} \sum_{i=1}^2 c_i(m_b) O_i(m_b) $, 
which includes both the {\it soft} (non-perturbative) contribution expressed by the 
operators $O_i(m_b)$ and the {\it hard} (perturbative) one represented by the Wilson 
coefficients $\ c_i(m_b)$. Both tree and penguin diagrams are computed and the final 
hadronic matrix element is given by:   
\begin{multline}
\mathcal{A}(\Lambda_b \to \Lambda V)= 
\frac{G_{F}}{\sqrt{2}} 
f_V E_V {\langle \Lambda | \bar{s} \Gamma_{\mu} b| \Lambda_b\rangle} \\ 
 \biggl\{ \mathcal{M}^T_{\Lambda_b}(\Lambda_b \to \Lambda V)-
\mathcal{M}^P_{\Lambda_b}(\Lambda_b \to \Lambda V) \biggr\}\ , \nonumber
\end{multline}
\noindent
with $\Gamma_{\mu} =\gamma_{\mu} (1-\gamma_{5})$, 
$\mathcal{M}^{T,P}_{\Lambda_b}(\Lambda_b \to \Lambda V)= V_{ckm}^{T,P} 
A^{T,P}_{V}(c_{i})$,  where $f_V$  is the vector-meson decay constant, $A^{T,P}_{V}(c_{i})$
are tree (T)  and penguin (P) amplitudes made of combinations of Wilson coefficients 
according to the nature of $V(1^-)$, $V_{ckm}$ are the 
Cabbibo-Kobayashi-Maskawa matrix elements and, finally the  ${\rho}^0-{\omega}$ mixing 
is taken into account for  the case $V  \to {\pi}^+ {\pi}^-$.  
%--------------------------------------------------------------------------------
\section{MAIN PHYSICAL RESULTS}
Several results which can be tested experimentally have been obtained. 
The most important ones are:
%-------------------------------------------------------------------------------
\subsection{ Branching ratios }
\noindent
In the framework of the factorization hypothesis, the color number is an 
effective parameter which is let {\it free}. The different branching ratios 
are proportionnal to the following width given by:
\begin{multline}
\Gamma(\Lambda_b \to \Lambda V)= \\
\frac{E_{\Lambda}+M_{\Lambda}}{M_{\Lambda_b}}\frac{P_V}{16\pi^2} 
 \int_{\Omega} |A_0(M_i)|^2 d\Omega\ , \nonumber 
\end{multline}
while the only experimental branching ratios (Ref.~\cite{Eidelman:2004wy}) is 
$\mathcal{BR}^{exp}(\Lambda_b \to \Lambda J/\psi)=(4.7\pm2.1 \pm 1.9)\times 10^{-4}$
 which permits to state that $2.0   \le  N^{eff}_c   \le  3.0.$ Numerical results are 
shown in Table 1.
%--------------------------------------------------------------------------------
\subsection{ Polarizations and asymmetries }
\noindent
Values of the resonance polarizations as well as their density matrix are essential 
ingredients for Monte-Carlo simulations, particularly for angular distributions.
They can be computed from our model as well as other parameters like $\Lambda$ 
helicity asymmetries. In Table 2 are listed the numerical results:
%%%%%%%%%%%%%%%%%%%%%%%%%%%%%%%%%%%%%%%%%%%%%%%%%%%%%%%%%%%%%%%%%%%%%%
 \begin{table}[hpt]
\caption{Polarizations and asymmetries in case of $\Lambda_b \to \Lambda V$ with 
$V$ is $\rho^0(\omega)$ or $J/\Psi.$}
\label{table:2}
 \renewcommand{\tabcolsep}{1.4pc} % enlarge column spacing
\renewcommand{\arraystretch}{1.1} % enlarge line spacing
\begin{tabular}{|l|c|c|}
 \hline
Parameter&$\Lambda {\rho}^0-{\omega}$&$\Lambda J/{\psi}$\\
\hline
${\alpha}^{\Lambda_b}_{AS}$&0.194&0.490\\
\hline
 ${\cal P}^{\Lambda}$&-0.21&-0.17\\
\hline
 ${\rho}^{\Lambda}_{+-}$&0.31&0.25\\
\hline
 ${\rho}^V_{00}$&0.79&0.66\\
\hline
\end{tabular}
\end{table}
%%%%%%%%%%%%%%%%%%%%%%%%%%%%%%%%%%%%%%%%%%%%%%%%%%%%%%%%%%%%%%%%%%%%%%
\noindent
It is worth noticing that 
(i) all these "geometrical" parameters do not depend on $N_c^{eff}$ but they are 
related directly to the weak decay process and, (ii) longitudinal polarizations 
of the vector mesons are {\it dominant}.
%--------------------------------------------------------------------------------
\subsection{Effects of ${\rho}^0 - {\omega}$ Mixing}
%%%%%%%%%%%%%%%%%%%%%%%% Histogramme de l'amplification de CP %%%%%%%%%%%%%%%%%%
\vskip -1.0cm
\begin{figure}[hp]
\begin{center}
\includegraphics*[width=0.9\columnwidth]{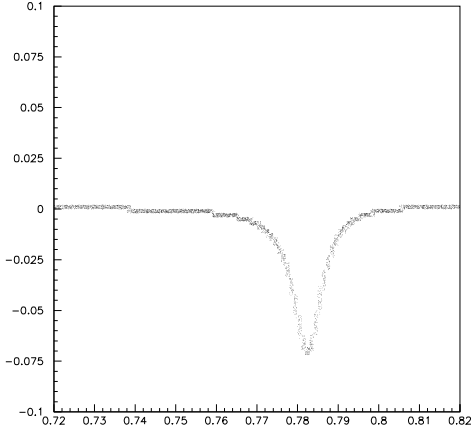}
\end{center}
\vskip -1.0cm
\caption{Branching ratio asymmetry, $a_{CP}(\omega)$, as a function of  the ${\pi}^+ {\pi}^-$ 
invariant mass in case of $\Lambda_b \to \Lambda \rho^0(\omega) \to \Lambda \pi^+\pi^-$ and 
for $N_c^{eff}=3$.}
\end{figure}
%%%%%%%%%%%%%%%%%%%%%%%%%%%%%%%%%%%%%%%%%%%%%%%%%%%%%%%%%%%%%%%%%%%%%%%%%%%%%%%%%
\noindent The asymmetry parameter between two conjugate channels defined by:
\begin{eqnarray}
a_{CP}(s_{\rho})=  \frac{\mathcal{BR}(\Lambda_b) - \mathcal{BR}(\bar \Lambda_b)}
{\mathcal{BR}(\Lambda_b)+\mathcal{BR}(\bar \Lambda_b)}\ , \nonumber  
\end{eqnarray}
where $ s_{\rho} \ \mathrm{is} \  {\pi}^+ {\pi}^- \ $ invariant mass, 
varies with $N_c^{eff}$ and it is usually too small, $a_{CP} \le 
 10^{-3}$. However in the case of ${\rho}^0 - {\omega} \to {\pi}^+ {\pi}^-$, this 
asymmetry is {\it amplified} in the vicinity of the  $\omega$ mass~\cite{Guo:2000uc}, as it is shown on 
Fig.~1, and it reaches $  7.5\% $ at the $\omega$ pole  for $ \ {N}^{eff}_c = 3.0$. This 
process is a new way to detect Direct CP Violation, as it has been already shown in 
beauty meson $B$ decays.
%-------------------------------------------------------------------------------
\section{TIME ODD OBSERVABLES}
The initial laboratory frame is transposed to the $\Lambda_b$ rest-frame as $(\vec {e_X},  \vec {e_Y}, 
\vec {e_Z})$ with  $\vec {e_Z}$ parallel to $\vec {e_3}$. For each resonance
with momentum $\vec p$,  a new frame is defined (Jackson~\cite{Jackson:1965}) as follows:
\begin{eqnarray}
\vec{e}_L =  \frac{\vec{p}}{p} , \    
 \vec{e}_T =  \frac{\vec{e}_Z \times \vec {e}_L}{|\vec{e}_Z \times \vec{e}_L|} \ ,   
 \vec{e}_N = \vec{e}_T \times \vec{e}_L\ . \nonumber
\end{eqnarray} 
%---------------------------------------------------------------------------------
\subsection{Vector-polarization}
\noindent
Each vector-polarization $\vec{\mathcal{P}}^{(i)}$  can be expanded on the new basis 
and its components are studied under Parity and Time-Reversal operations.
\begin{eqnarray}
\vec{\mathcal{P}}^{(i)} = P_L^{(i)} \vec{e}_L + P_N^{(i)} \vec{e}_N + P_T^{(i)} \vec{e}_T\ , \nonumber
\end{eqnarray}
with  
$P_j^{(i)} = \vec{\mathcal{P}}^{(i)} \cdot \vec {e}_j \  \mathrm{and} \ j = L,N,T $. 
%%%%%%%%%%%%%%%%%%%%%%%%%%%%%%%%%%%%%%%%%%%%%%%%%%%%%%%%%%%%%%%%%%%%%%%%%%%%%%%%
\begin{table}[hbp]
\caption{Vector-polarization under Parity and TR operations.}
\label{table:3}
 \renewcommand{\tabcolsep}{1.4pc} % enlarge column spacing
\renewcommand{\arraystretch}{1.1} % enlarge line spacing
\begin{tabular}{|c|c|c|}
\hline
Observable&Parity&TR\\
\hline
$\vec s$&Even&Odd\\
$\vec {\cal P}$&Even&Odd\\
\hline
$\vec {e_Z}$&Even&Even\\
$\vec {e_L}$&Odd&Odd\\
$\vec {e_T}$&Odd&Odd\\
$\vec {e_N}$&Even&Even\\
\hline
${P_L}$&Odd&Even\\
${P_T}$&Odd&Even\\
${P_N}$&Even& ODD\\
\hline
\end{tabular}
\end{table}
%%%%%%%%%%%%%%%%%%%%%%%%%%%%%%%%%%%%%%%%%%%%%%%%%%%%%%%%%%%%%%%%%%%%%%%%%%%%%%%%%%
\noindent
It could be noticed that if the normal component  $P_N$ is non equal to zero, 
it would be a clear signal of   TR Violation. See Table 3 for TR and parity operations on 
vector polarization.
%-----------------------------------------------------------------------------------
\subsection{Special angles}
\noindent
${\vec n}_{\Lambda}$  and   $\vec{n}_V$ are defined respectively as the 
unit normal vetors to $\Lambda$ and $V$ decay planes. 
$\vec{n}_{\Lambda} = \frac{\vec{p}_p \times \vec{p}_{\pi}}{|\vec{p}_p \times \vec{p}_{\pi}|}\ , \ \
\vec{n}_V = \frac{ \vec{p}_{l^+} \times  \vec{p}_{l^-}}{ | \vec{p}_{l^+} \times \vec{p}_{l^-}|}\ , \ \
 \mathrm{or} \ \  \vec{n}_V = \frac{ \vec{p}_{h^+} \times  \vec{p}_{h^-}}{ |\vec{p}_{h^+} \times 
\vec{p}_{h^-}|}\ $. 
Those vectors are even under TR; but, if we compute the cosine and the sine of 
their azimutal angles, ${\phi}_{(n_i)}= {\phi}_{ {\vec n}_{\Lambda}}\  , 
\ {\phi}_{{\vec n}_V}$, as it was suggested by Seghal~\cite{Sehgal:1999vg} and 
Wolfenstein~\cite{Wolfenstein:1999xb} for the decay $K^0_L \to {\pi}^+ {\pi}^- e^+ e^-$,  
$
\vec{u}_i = \frac{ \vec{e}_Z \times \vec{n}_i}{ |\vec{e_Z} \times \vec{n}_i|}\ , \ \  \cos
{\phi}_{(n_i)} = \vec{e}_Y \cdot \vec{u}_i\ , \ \  \sin {\phi}_{(n_i)} = 
\vec{e}_Z \cdot  (\vec{e}_Y \times \vec{u}_i)\ , 
$
we notice that	$\cos{\phi}_{(n_i)}$ and $\sin {\phi}_{(n_i)}$  are odd under 
TR; and these asymmetries depend essentially on the azimuthal angle distributions 
of the $\Lambda$  resonance in the $\Lambda_b$ rest-frame, which analytical 
expression is given by:
\begin{multline}
{d\sigma}/{d\phi}   \propto 1+  \\
  \!\! \frac{\pi}{2} {\alpha_{AS}}
\Big({\Re e({\rho}_{+-}^{\Lambda_b}) \cos \phi - \Im m({\rho}_{+-}^{\Lambda_b}) 
\sin \phi } \Big)\ . \nonumber
\end{multline}
\newline
By choosing conservative values for the non-diagonal elements of the 
$\Lambda_b$ polarization density matrix: $  {\Re e({\rho}_{+-}^{\Lambda_b})}  
=  -{\Im m({\rho}_{+-}^{\Lambda_b})} =  {\sqrt{2}}/2$,
the following asymmetries are obtained:
\begin{itemize}
\item{For $\Lambda_b \to \Lambda J/{\psi}$, one obtains:
\newline
$AS(\cos {\phi}_{{\vec n}_{\Lambda}}) = 4.3\%\ ,$ \\
$AS(\sin{\phi}_{{\vec n}_{\Lambda}}) = -5.5\%\ .$}
\item{For $\Lambda_b \to \Lambda {\rho}^0(\omega)$, one obtains:
\newline
$AS(\cos {\phi}_{{\vec n}_{\Lambda}}) = 2.4\%\ ,$ \\
$AS(\sin{\phi}_{{\vec n}_{\Lambda}}) = -2.7\%\ .$}
\end{itemize}
\noindent But, no asymmetries in  $\cos {\phi}_{{\vec n}_V}$ and  $\sin {\phi}_{{\vec n}_V}$ 
of the resonances $V(1^-)$  are seen. A realistic explanation for these different 
asymmetries is suggested: (i) T-Odd or  
TRV effects  appear in processes where already Parity is violated like   
$\Lambda \to p {\pi}^- $, (ii) while for processes as $V(1^-) \to {\ell}^+ {\ell}^-, 
h^+ h^-$ where Parity is conserved, T-Odd effects are absent.
%----------------------------------------------------------------------------------  
\section{CONCLUSION}
Complete calculations of the processes $\Lambda_b \to \Lambda V(1^-)$ 
have been performed. On the kinematics side, they are based on the helicity 
formalism and the use of resonance polarization density matrices. On the dynamics 
side, sophisticated methods using the OPE formalism and HQET have been developed 
in the framework of the {\it factorization hypothesis}. Our model is entirely built 
in the framework of the Standard Model. No ingredients coming from "Beyond Standard Model" 
are introduced.  The only unknown parameters are the elements of the polarization density matrix of 
the produced $\Lambda_b$  in p-p collisions. By adopting conservative values for these 
matrix elements and an entirely polarized  $\Lambda_b$, asymmetries related to Time-Reversal 
Invariance are observed. So, crucial questions may arise: (i) Is there {\it any dynamics} behind Time-Reversal 
Violation? (ii) Is this dynamics related to the CKM mechanism ? Whatever are the answers, it is plausible 
to assert that Time-Reversal Violation or T-odd processes 
are real challenges for the next LHC experiments. 
%-----------------------------------------------------------------------------------

\subsection*{Acknowledgements}
\noindent
One of us (Z.J.A.) is very grateful to Pr Stefan Narison, chairman of the QCD06 International Conference, for his warm
hospitality at Montpellier city and for all the illuminating discussions he got with him.

%--------------------------------------------------------------------------------
\end{document}